# Record-proof optical communication enabled by a quantum receiver with lossless optical decryption

Ken Tanizawa*, Takayuki Kurosu, Fumio Futami, Kenji Nakahira, Kentaro Kato

Quantum ICT Research Institute, Tamagawa University, 6-1-1 Tamagawagakuen, Machida, Tokyo, Japan.

*Correspondence: Ken Tanizawa (tanizawa@lab.tamagawa.ac.jp)

## Abstract

We propose record-proof physical-layer encryption (PLE) based on quantum-optimal reception. Unlike conventional symmetric-key encryption algorithms, the proposed scheme prevents decryption of recorded signals even if the secret key is disclosed after transmission. Existing PLE schemes rely on practical limitations of the eavesdropper's receiver in pursuing record-proof security; ours instead exploits the noncommutativity of conjugate optical quadratures to give the legitimate receiver an intrinsic sensitivity advantage. A phase-controlled Kennedy receiver performs lossless optical decryption and near-optimal binary-state discrimination. We demonstrate record-proof security with a 3.0-dB channel-loss budget, showing that quantum receivers offer cryptographic functions beyond sensitivity enhancement.

## 1. Introduction

Secure communication generally involves authentication, key establishment, and message encryption. Typically, optical communication systems employ symmetric-key encryption algorithms such as the Advanced Encryption Standard (AES) for message encryption [1]. The security of AES is solely based on computational complexity, and the so-called harvest now, decrypt later (HNDL) attack is an inherent threat. The encrypted data carry information not only about messages but also about the secret key. Thus, the possibility of eventually deducing the key from the stored encrypted data through future cryptoanalysis cannot be eliminated. This risk can be mitigated by frequently updating the key using quantum key distribution (QKD) [2].

Physical-layer encryption (PLE) addresses the HNDL threat from a different perspective: as a signal-level symmetric-key encryption, it combines algorithms and physics to prevent effective harvesting. The quantum noise stream cipher, originally proposed as αη [3] and Y-00 protocol [4], utilizes high-order modulation with a pre-shared secret key, and quantum uncertainty inevitably induces errors in signals received by eavesdroppers without the secret key. Data demodulated from the received signals contain numerous bit errors, impeding subsequent offline cryptoanalysis in HNDL attacks. Furthermore, recent advances in coherent detection and digital signal processing have enhanced the practicality of quantum noise stream cipher systems in optical communications [5,6].

A more advanced level of security, known as record-proof security, was proposed in the context of PLE. Record-proof security ensures that encrypted data remain undecipherable even if an eavesdropper intercepts and stores all signals accessible at the receiver and subsequently obtains the secret key. Unlike conventional symmetric-key encryption schemes, including the one-time pad, whose ciphertext can be fully decrypted once the key is revealed, record-proof security provides long-term confidentiality without requiring strict key disposal, and fundamentally eliminates the threat of HNDL attacks. One practical approach for achieving record-proof security is the combination of optical spectrum spreading and phase encoding with a secret key [7–9]. This technique enables a negative optical signal-to-noise ratio transmission and can realize record-proof security. However, the security of PLE based on optical spectrum spreading relies on the practical assumption that an eavesdropper does not have access to an ultrawideband receiver capable of capturing the entire spread spectrum.

In this paper, we demonstrate record-proof security without practical restrictions on an eavesdropper's receiver in HNDL attacks. The invariant advantage of a legitimate receiver with a secret key is generated based on the difference in receiver sensitivity derived from the quantum incompatibility of the conjugate quadratures of light. A technical issue in implementing this system is the signal attenuation caused by optical phase manipulation for decryption. We propose a near-optimal quantum receiver that incorporates lossless optical decryption for legitimate reception. Experimental results show that the receiver achieves record-proof security with a channel-loss budget of 3.0 dB. The present work demonstrates, for the first time, that the sensitivity advantage arising from quantum-optimal reception can be converted into record-proof security, thereby realizing security inaccessible to conventional cryptographies.

## 2. Operating principles

For record-proof security without practical restrictions, a legitimate receiver uses a pre-shared secret key to achieve better sensitivity than the eavesdropper. This is realized by forcing the eavesdropper to simultaneously measure the in-phase and quadrature-phase (I and Q) components, whereas a legitimate receiver can detect the I or Q components for the demodulation of message bits. The I and Q components, corresponding to the two conjugate quadratures of light, are noncommutative observables, which is a fundamental principle of quantum mechanics. In contrast to orthogonal physical degrees of freedom, such as polarization and wavelength, the eavesdropper's sensitivity disadvantage induced by *simultaneous* I and Q reception is fundamental and persists regardless of the practical constraints. This concept enables record-proof security without imposing practical restrictions on the eavesdropper's receiver.

The configuration for implementing the proposed system is shown in Fig. 1. The coherent light from the laser is modulated, and high-order encrypted phase-shift keying (PSK) signals are generated based on the Y-00 protocol with a 256-bit pre-shared secret key (see Section 1 of Supplement 1). In the protocol, the message bits are first modulated as a binary PSK (BPSK) signal $|\pm\alpha\rangle$. The phase of the BPSK signal is then randomized symbol-by-symbol according to the key information. The phase-randomized encrypted signal is expressed as $|\pm\alpha e^{i\theta_{enc}(k)}\rangle$, where $\theta_{\mathrm{enc}}(k)$ is the phase of the $k$-th symbol for encryption. The resolution of the phase randomization is set sufficiently fine, corresponding, for example, to a $2^{10}$-PSK modulation format, such that quantum noise masks adjacent signals, thereby protecting the secret key during encrypted communication. The encrypted high-order signals propagate through the fiber link. On the side of the legitimate receiver with the secret key, an inverse phase shift is applied to the signal to recover the BPSK signals that carry the message bits. Subsequently, an optimal or near-optimal quantum receiver for binary-state discrimination [10–15] is employed to enhance the receiver sensitivity beyond that achievable with conventional classical detection.

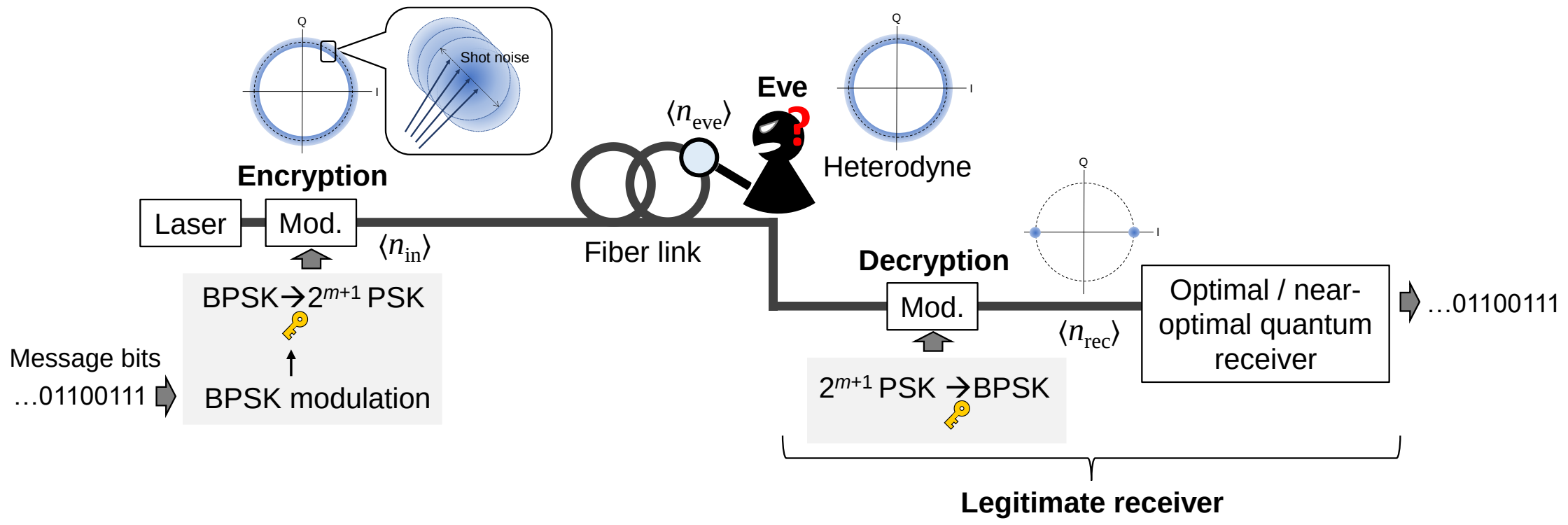

**Figure 1**: Proposed record-proof encrypted communication system using optimal/near-optimal quantum receiver.

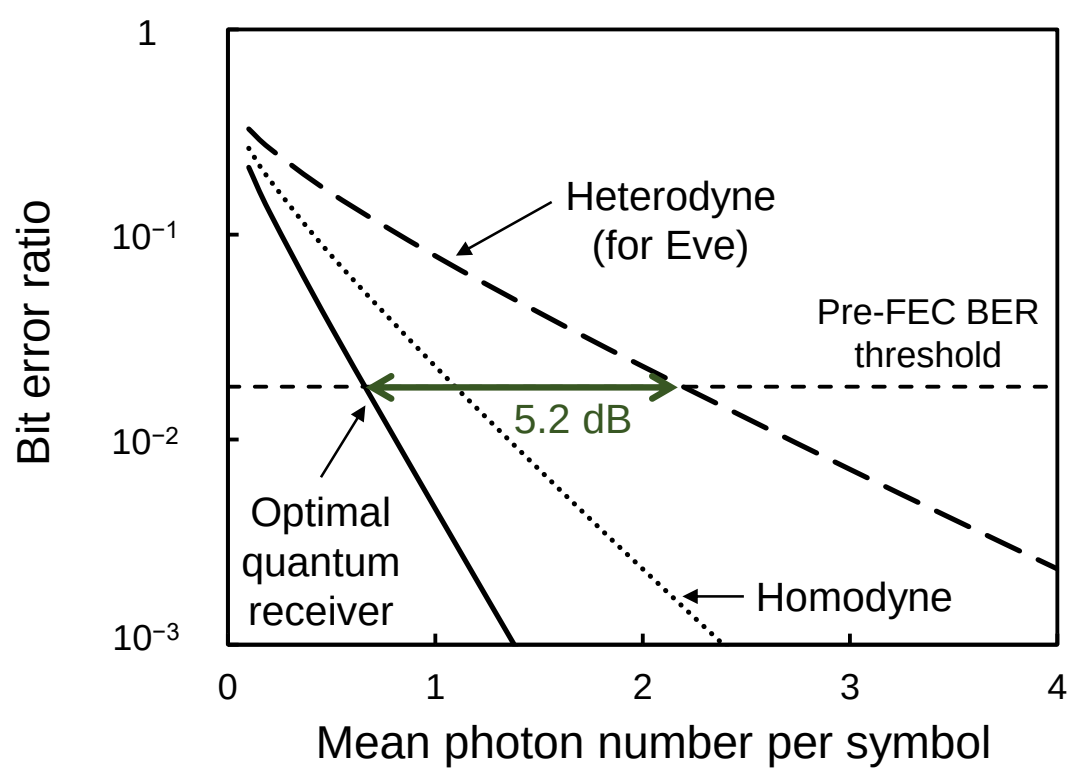

**Figure 2**: Theoretical bit error ratios (BERs) for heterodyne detection (Eve), homodyne detection, and the optimal quantum reception. At pre-forward error correction BER threshold of $1.81 \times 10^{-2}$, the sensitivity difference between the optimal quantum and heterodyne receivers gives a 5.2-dB channel-loss budget.

By contrast, an eavesdropper without a secret key cannot recover the original BPSK signal required for binary state discrimination. Instead, to perform an HNDL attack, the eavesdropper must acquire the complete complex-field information of the encrypted phase-randomized signal at the time of interception, so that any future key disclosure can be exploited offline. Therefore, we consider an ideal shot-noise-limited heterodyne receiver that provides the optimum classical strategy for simultaneously acquiring both the I and Q components. Even ideal amplification cannot improve the eavesdropper's reception performance, as simultaneous amplification of I and Q components has a quantum-limited noise figure of 3 dB [16], equivalent in noise performance to ideal heterodyne detection. No additional implementation constraints are imposed on the eavesdropper's receiver.

The system requires appropriate signal power settings for record-proof security. The signal power launched into a link, represented as the mean photon number per symbol $\langle n_{\mathrm{in}}\rangle$, is set less than $\langle n_{\mathrm{hetero}}\rangle$, where $\langle n_{\mathrm{hetero}}\rangle$ is the minimum mean photon number required for ideal heterodyne reception at the pre-forward error correction (pre-FEC) bit error ratio (BER) threshold. A classical eavesdropper obtains a photon number $\langle n_{\mathrm{eve}}\rangle$; however, $\langle n_{\mathrm{eve}}\rangle$ is smaller than $\langle n_{\mathrm{hetero}}\rangle$, and the reception can never reach the pre-FEC BER threshold. Meanwhile, the legitimate quantum receiver for binary discrimination has better sensitivity: the required photon number $\langle n_{\mathrm{oqr}}\rangle$ is smaller than $\langle n_{\mathrm{hetero}}\rangle$. Thus, provided that the channel loss is lower than a certain value, the received photon number $\langle n_{\mathrm{rec}}\rangle$ can be larger than $\langle n_{\mathrm{oqr}}\rangle$, and the pre-FEC BER threshold is satisfied.

We estimate the achievable channel-loss budget by assuming optimum quantum discrimination of the recovered BPSK signals for the legitimate receiver and ideal shot-noise-limited heterodyne reception followed by offline decryption using the correct key for the eavesdropper. The performance of the eavesdropper is equivalent to that of the BPSK reception with ideal heterodyne detection. Fig. 2 compares the theoretical BER curves of the BPSK reception using an optimum quantum receiver and a heterodyne receiver with unit quantum efficiency (Supplement 1, Section 2). The curve for the optimal quantum receiver corresponds to the Helström bound, which represents the quantum limit of binary-state discrimination [17]. The pre-FEC BER threshold is set to $1.81\times10^{-2}$, a typical value for current optical coherent transceivers [18]. To achieve record-proof security, the input mean photon number $\langle n_{\mathrm{in}}\rangle$ should satisfy $\langle n_{\mathrm{in}}\rangle\leq\langle n_{\mathrm{hetero}}\rangle = 2.2$. Under this condition, the available channel-loss budget is 5.2 dB, which corresponds to the difference in sensitivity between the legitimate quantum receiver and heterodyne receiver (Eve). In this system, the decryption, which recovers BPSK signals in the optical domain, enables the use of a highly sensitive quantum receiver. However, conventional decryption requires direct manipulation of the optical signal using an optical modulator, which inevitably introduces insertion loss of several decibels, thereby significantly reducing the loss budget in the proposed system.

We propose a novel static displacement quantum receiver with local state phase modulation for lossless decryption. Here we consider the simplest displacement with a local state $|\beta\rangle = |\alpha\rangle$, which is known as the Kennedy receiver [10,15]. The concept of local phase modulation for decryption is applicable to other static and adaptive quantum receivers [19]. The proposed Kennedy receiver is shown in Fig. 3. The input encrypted signal is expressed as $|\pm\alpha e^{i\theta_{enc}(k)}\rangle$. Instead of directly modulating the signal, a phase modulator (PM) shifts the local state phase by $\theta_{\mathrm{dec}}(k)$ for decryption. The local state is expressed as $|\alpha e^{i\theta_{dec}(k)}\rangle$, and $\left|\pm\alpha e^{i\theta_{enc}(k)} + \alpha e^{i\theta_{dec}(k)}\right\rangle = e^{i\theta_{dec}(k)\hat{n}}\left|\pm\alpha e^{i(\theta_{enc}(k)-\theta_{dec}(k))} + \alpha\right\rangle$ is obtained through the displacement operation. A legitimate receiver sets $\theta_{\mathrm{dec}}(k)$ to $\theta_{\mathrm{enc}}(k)$ according to the pre-shared secret key, thereby canceling the encryption phase shift. No signal photons are lost during this process; thus, lossless decryption is achieved. Finally, the receiver discriminates between $|0\rangle$ and $e^{i\theta_{dec}(k)\hat{n}}|2\alpha\rangle$, corresponding to the BPSK phases 0 and π, according to whether the single photon detector (SPD) registers a click, as in the conventional Kennedy receiver. Fig. 4 compares the theoretical BER curves of the BPSK reception using an optimum quantum receiver, Kennedy receiver, and a heterodyne receiver. The proposed Kennedy receiver incorporating lossless decryption achieves performance equivalent to that of the conventional Kennedy receiver. Thus, the channel-loss budget achievable with the proposed Kennedy receiver was estimated to be 4.2 dB.

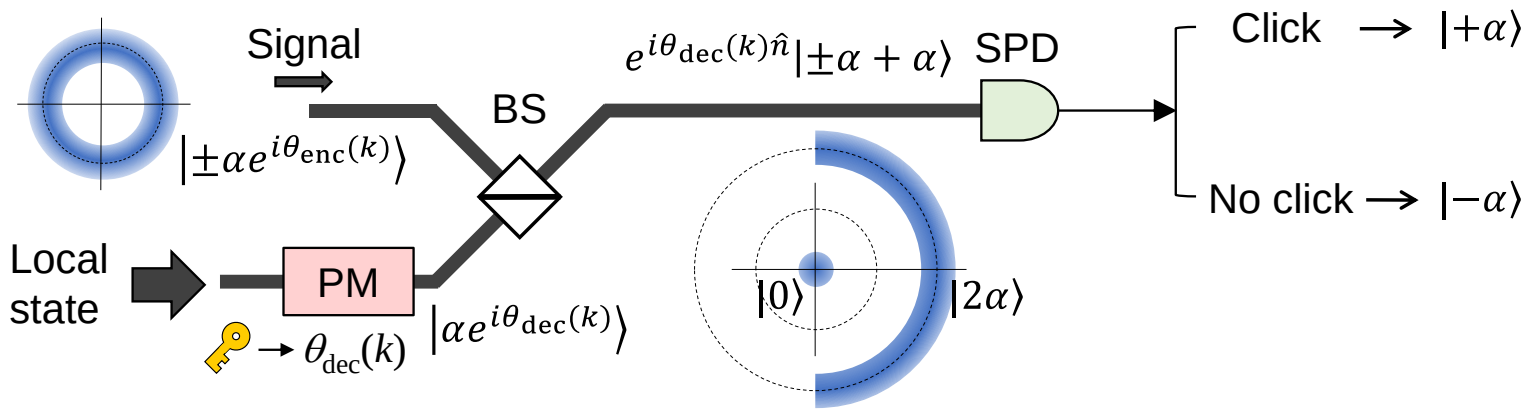

**Figure 3**: Proposed Kennedy receiver with lossless decryption. PM, phase modulator; BS, beam splitter; SPD, single-photon detector. The local state phase is modulated using the secret key information for lossless decryption.

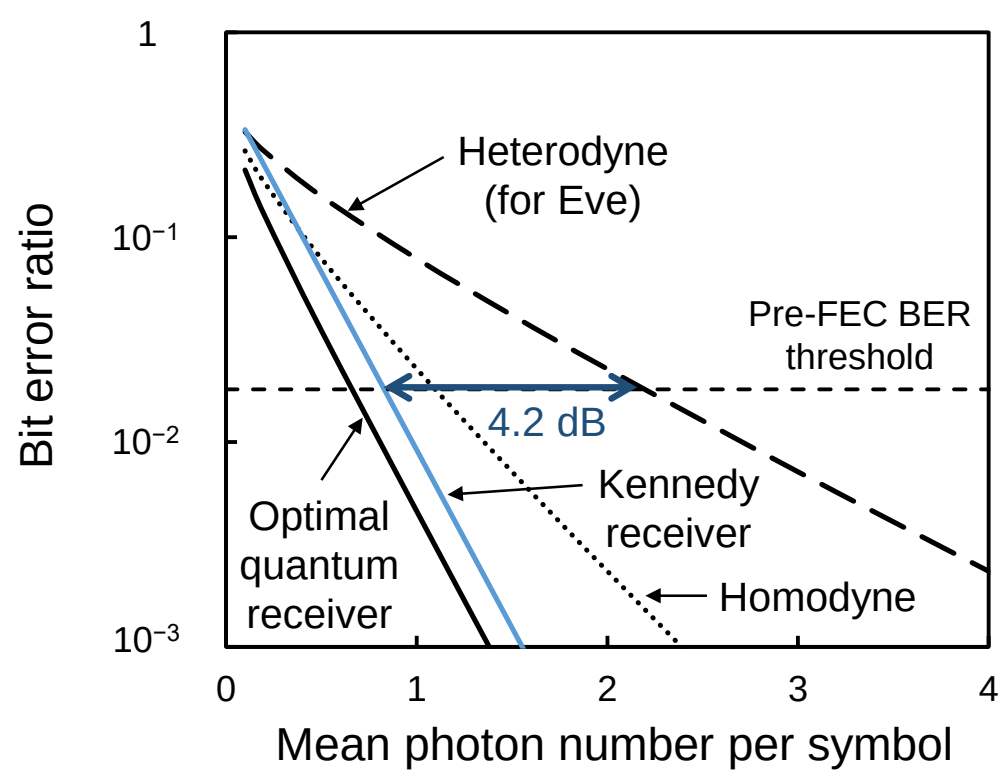

**Figure 4**: Theoretical BERs for heterodyne detection (Eve), homodyne detection, Kennedy receiver, and optimal quantum, receiver. At a pre-FEC BER threshold of $1.81 \times 10^{-2}$, the channel-loss budget of Kennedy receiver is 4.2 dB.

## 3. Experiments

We demonstrated record-proof security using a Kennedy receiver with lossless decryption in a fiber-based configuration as shown in Fig. 5. A continuous-wave laser operating at a wavelength of 1550.12 nm (NKT Photonics, Koheras ADJUSTIK C15) served as the common optical source for the signal, local-state, and reference paths. The output light was routed by an optical switch (OSW) driven by a square-wave control signal generated by an arbitrary waveform generator (AWG) (Zurich Instruments, HDAWG). The frequency and duty cycle were 36.621 kHz and 50 %, respectively. The switch alternately generated communication and reference intervals, enabling encrypted signal transmission and phase stabilization using the same optical source.

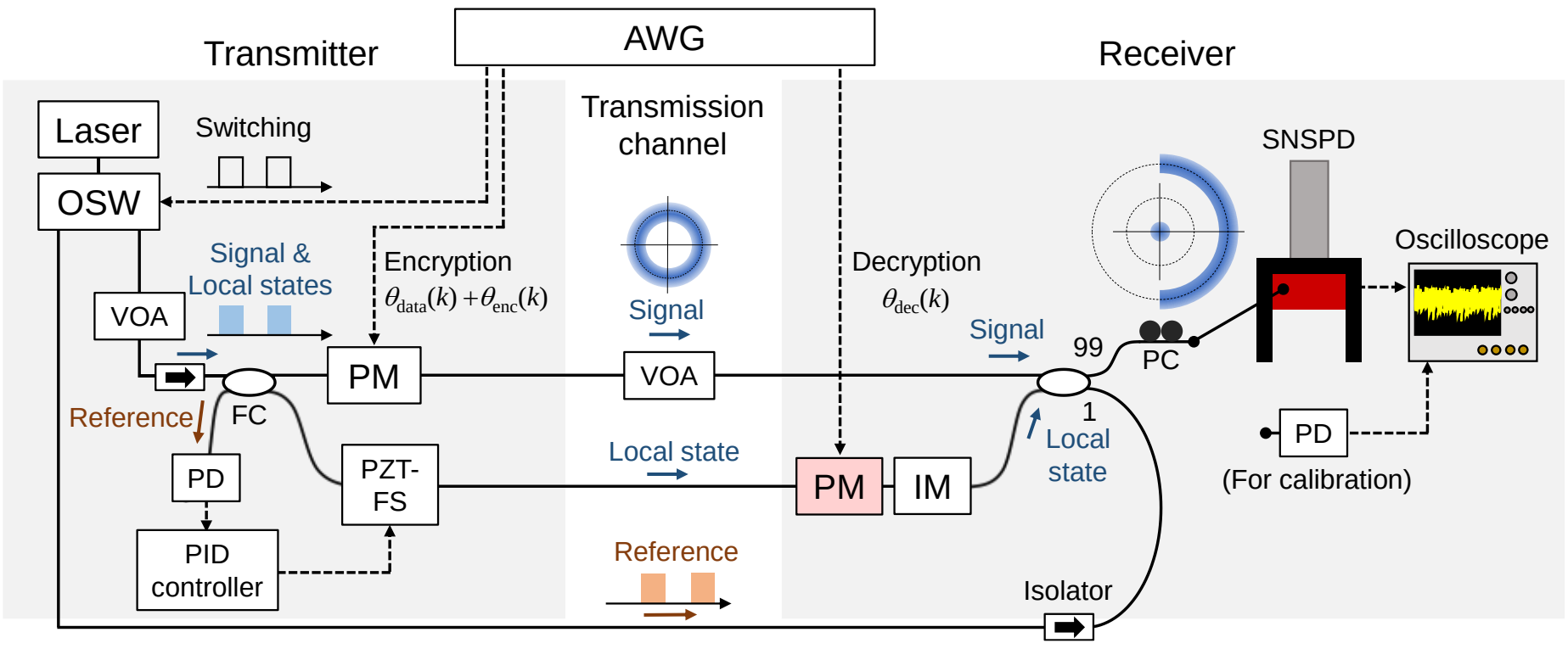

**Figure 5**: Schematic of the experimental setup for record-proof communication using a Kennedy receiver with lossless decryption. AWG, Arbitrary waveform generator; OSW, optical switch; VOA, variable optical attenuator; FC, fiber coupler; PM, phase modulator; PD, photodetector; PZT-FS, piezoelectric fiber stretcher; IM, intensity modulator; PC, polarization controller; SNSPD, superconducting nanowire single photon detector.

The optical field was divided into signal and local-state paths by a fiber coupler (FC). During the communication interval, the signal phase was modulated using a phase modulator (PM) driven by the AWG. The modulation signal consisted of the sum of the BPSK data phase, $\theta_{data}(k)$, and the encryption phase, $\theta_{enc}(k)$, where $\theta_{data}(k) \in \{0, \pi\}$ and $-\pi/2 \leq \theta_{enc}(k) < \pi/2$. The encryption phase can take one of $2^9$ discrete levels according to the Y-00 protocol, thereby generating the encrypted $2^{10}$ (=1024) PSK signal. The transmission channel was emulated using a variable optical attenuator (VOA). The local-state path was optically isolated from the signal path and included a piezoelectric fiber stretcher (PZT-FS) for phase stabilization. During the reference interval, a fraction of the optical field propagated backward through the interferometer and was detected by a photodiode (PD). The detected signal was processed by a proportional-integral-derivative (PID) controller (New Focus, LB1005), which drove the PZT-FS to compensate for environmental phase fluctuations and maintain a stable relative phase between the signal and local-state paths.

At the receiver, optical decryption was performed by applying the phase shift $\theta_{dec}(k) = \theta_{enc}(k)$ to the local state using a second PM. The decryption phase was generated by the same AWG according to the pre-shared secret key. Then, the local-state amplitude was adjusted using an intensity modulator (IM) and combined with the received signal at a 99:1 fiber coupler to simultaneously implement the phase shift for decryption and the displacement operation required for Kennedy reception. Because the phase shift was applied to the local state rather than the encrypted signal itself, decryption was achieved without introducing signal attenuation. The displaced optical state was detected using a superconducting nanowire SPD (SNSPD) (Single Quantum, Single Quantum Eos) with a system detection efficiency of 91 %. A polarization controller (PC) was inserted before detection to maximize the receiver sensitivity. The electrical output of the SNSPD was recorded by a real-time oscilloscope (Tektronix, MSO64), and binary state discrimination was performed based on the detector click events. Prior to the measurements with the SNSPD, the output optical power was monitored using an auxiliary photodiode located at the receiver side, and the local-state amplitude was adjusted to maximize the extinction ratio of the displacement interference. The BER characteristics were obtained offline by comparing the recovered data sequence with the transmitted pseudo-random bit sequence.

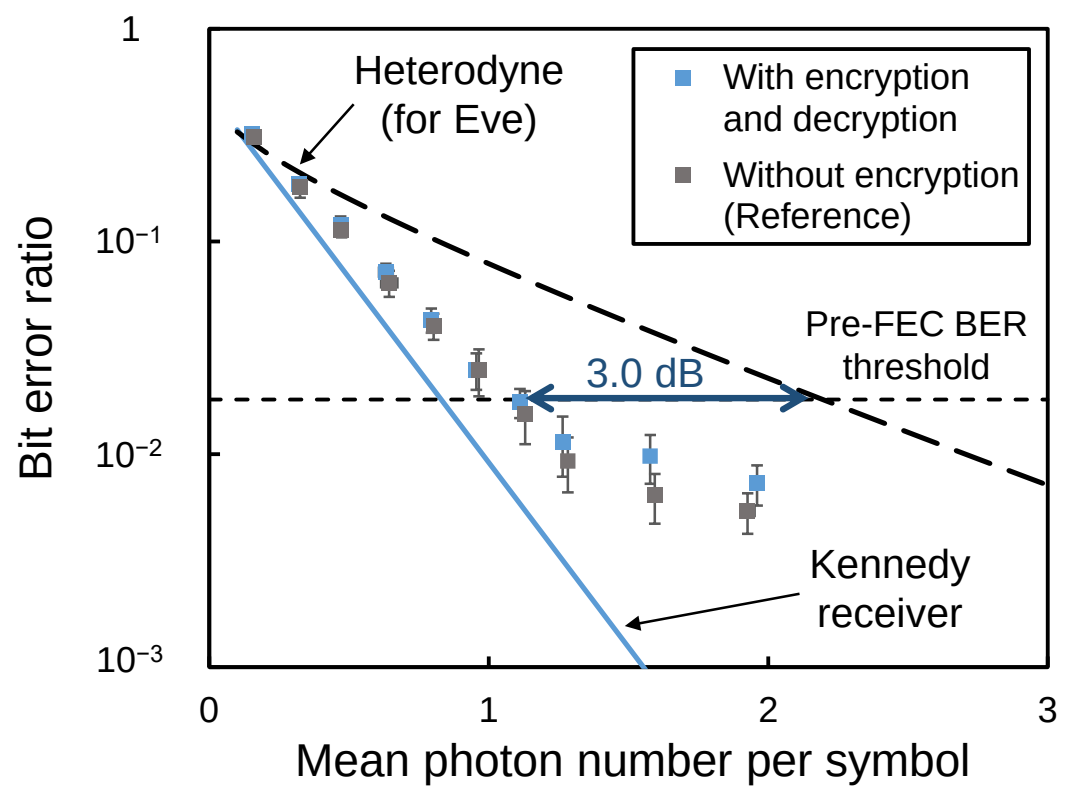

**Figure 6**: Experimental BER characteristics. The blue and black plots indicate decrypted BPSK signal and reference BPSK signal, respectively. The channel-loss budget for record-proof security is 3.0 dB at a pre-FEC BER threshold of $1.81 \times 10^{-2}$.

Fig. 6 compares the measured BERs of the encrypted/decrypted and reference BPSK signals with the theoretical curves. The mean photon number takes into account the losses of the FC and PC, as well as the detection efficiency of the SNSPD. The BER of the decrypted signal closely followed that of the reference BPSK signal, indicating successful decryption without signal loss due to the decryption. The slight BER penalty observed in the encrypted/decrypted case may originate from the residual phase errors caused by imperfect cancellation between the encryption and decryption phase modulations. In both cases, an error floor appeared when the mean photon number exceeded approximately 1 per symbol. This behavior is mainly caused by the limited extinction ratio of the interference between the signal and the local states. As a result, a nonzero residual optical field remains even when the displaced state should ideally be in a vacuum state, leading to false click events and, consequently, detection errors. At the pre-FEC BER threshold of $1.81 \times 10^{-2}$, the measured channel-loss budget was 3.0 dB. Considering the recent advancements in low-loss hollow-core fiber technology with an attenuation below 0.1 dB/km [20], the system can potentially achieve fiber transmission over 30 km. The observed performance was approximately 1.2 dB below the theoretical limit of the Kennedy receiver. This degradation is primarily attributed to receiver-side photon losses, including the finite detection efficiency of the SNSPD and excess losses in the FC and PC. Therefore, the proposed receiver can achieve improved performance by adopting lower-loss and higher-efficiency devices.

Record-proof security is guaranteed at the instant legitimate reception is completed, provided that the eavesdropper does not possess a secret key. To satisfy this requirement, we employed the Y-00 protocol to prevent key leakage during encrypted communications. The encrypted PSK used $2^{10}$ phase levels. At a channel-input mean photon number $\langle n_{\mathrm{in}} \rangle$ of 2.2, which satisfies the requirement for the record-proof security, approximately 162 phase levels are masked by the quantum noise (see Section 3 of Supplement 1). This quantum noise masking effect inevitably induces detection errors in eavesdropper's illegitimate reception. A symbol error ratio for an eavesdropper is higher than 0.995, even if all photons are captured using an ideal heterodyne receiver. Therefore, the encrypted phase levels cannot be correctly recorded, thereby preventing the recovery of the secret key through subsequent offline cryptanalysis. Furthermore, periodic key updates, such as those enabled by QKD, can significantly reduce the risk of key leakage during encrypted communication by shortening the time window available for offline cryptanalysis, thus supporting record-proof security.

## 4. Conclusion

We experimentally demonstrated record-proof secure communication using a Kennedy receiver with lossless optical decryption. The proposed system achieved a channel loss budget of up to 3.0 dB while maintaining record-proof security. Although the bit rate was not optimized in this experiment, the achievable rate is limited by the SPD performance. Further advances in SPD technology are essential for extending our approach to high-speed optical communications. To the best of our knowledge, this is the first experimental demonstration of record-proof security without any practical constraints on the eavesdropper's receiver in HNDL attacks. Furthermore, the work extends the advantage of quantum receivers from sensitivity enhancement to cryptographic functionality beyond conventional symmetric-key encryption.

## Conflict of Interest

The authors declare no conflicts of interest.

## Funding

This work was partly supported by the Japan Society for the Promotion of Science (JSPS) KAKENHI under Grant JP25K01233 and the Air Force Office of Scientific Research under award number FA2386-22-1-4030.

## References

1. Ciena Application Note, 'WaveLogic Encryption Solution,' https://www.ciena.com/insights/app-notes/WaveLogic-Encryption-Solution-prx.html, accessed on 7 Aug 2026.
2. M. Pistoia, O. Amer, M. R. Behera, J. A. Dolphin, J. F. Dynes, B. John, P. A. Haigh, Y. Kawakura, D. H. Kramer, J. Lyon, N. Moazzami, T. D. Movva, A. Polychroniadou, S. Shetty, G. Sysak, F. Toudeh-Fallah, S. Upadhyay, R. I. Woodward, A. J. Shields "Paving the Way towards 800 Gbps Quantum-Secured Optical Channel Deployment in Mission-Critical Environments," arXiv [quantum-ph], arXiv:2202.07764v2.
3. G. Barbosa, E. Corndorf, P. Kumar, and H. P. Yuen, "Secure communication using mesoscopic coherent states," Phys. Rev. Lett., 90, 227901 (2003).
4. O. Hirota, M. Sohma, M. Fuse, and K. Kato, "Quantum stream cipher by Yuen 2000 protocol: Design and experiment by intensity modulation Scheme," Phys. Rev. A, 72, 022335 (2005).
5. X. Chen, K. Tanizawa, P. Winzer, et al., "Experimental demonstration of 4,294,967,296-QAM based Y-00 quantum stream cipher template carrying 160-Gb/s 16-QAM signals," Opt. Express, 29(4), 5658 (2021).
6. K. Tanizawa and F. Futami, "Ultra-long-haul digital coherent PSK Y-00 quantum stream cipher transmission system," Opt. Express, 29(7), 10451 (2021).
7. E. Wohlgemuth, Y. Yoffe, P. N. Goki, M. Imran, F. Fresi, P. D. Lakshmijayasimha, R. Cohen, P. Anandarajah, L. Pot`ı, and D. Sadot, "Stealth and secured optical coherent transmission using a gain switched frequency comb and multi-homodyne coherent detection,” Opt. Express, 29(24), 40462 (2021).
8. E. Wohlgemuth, Y. Yoffe, I. Attia, M. Imran, P. D. Lakshmijayasimha, A. Marotta, P. N. Goki, H. Shalom, F. Fresi, C. Antonelli, A. Mecozzi, P. Anandarajah, L. Pot`ı, and D. Sadot, “A field trial of multi-homodyne coherent detection over multi-core fiber for encryption and steganography,” J. Lightwave Technol. 41(9), 2723 (2023).
9. R. Cohen, E. Wohlgemuth, Y. Yoffe, Y. Yalinevich, I. Attia, A. Yalinevich, R. Yehoash, A. Rabinovich, D. Sadot, "Cryptanalysis of Practical Optical Layer Security Based on Phase Masking of Mode-Locked Lasers and Multi-Homodyne Coherent Detection," J. Lightwave Technol. 42(19), 6712 (2024).
10. R. S. Kennedy, “A near-optimum receiver for the binary coherent state quantum channel,” Research Laboratory of Electronics, MIT, Quarterly Progress Report, 108, 219 (1973).
11. S. J. Dolinar, “An optimum receiver for the binary coherent state quantum channel,” Research Laboratory of Electronics, MIT, Quarterly Progress Report, 111, 115 (1973).
12. C. Wittmann, M. Takeoka, K. N. Cassemiro, M. Sasaki, G. Leuchs, and U. Anderson, “Demonstration of Near-Optimal Discrimination of Optical Coherent States”, Phys. Rev. Lett., 101, 210501 (2008).
13. K. Tsujino, D. Fukuda, G. Fujii, et al., “Quantum Receiver beyond the Standard Quantum Limit of Coherent Optical Communication,” Phys. Rev. Lett., 106, 250503 (2011).
14. M. T. DiMario and F. E. Becerra, “Robust Measurement for the Discrimination of Binary Coherent States,” Phys. Rev. Lett., 121, 023603 (2018).
15. M.L. Shcherbatenko, M.S. Elezov, D.V. Sych, G.N. Goltsman, "Sub-shot-noise-limited fiber-optic quantum receiver," Physical Review A, 101(3), 032306 (2020).
16. E. Desurvire, “On the Physical Origin of the 3-dB Noise Figure Limit in Laser and Parametric Optical Amplifiers,” Optical Fiber Technol. 5(1), 40–61 (1999).
17. C. W. Helstrom, “Quantum Detection and Estimation Theory,” Academic Press, New York, 1976.
18. W. Wang, Z. Long, W. Qian, K. Tao, Z. Wei, S. Zhang, Z. Feng, Y. Xia, and Y. Chen, "Real-Time FPGA Investigation of Potential FEC Schemes for 800G-ZR/ZR+ Forward Error Correction," J. Lightwave Technol. 41(3), 926 (2023).
19. I. A. Burenkov, M. V. Jabir, and S. V. Polyakov, “Practical quantum-enhanced receivers for classical communication,” AVS Quantum Sci., 3, 025301 (2020).
20. M. Petrovich, E. N. Fokoua, Y. Chen, et al., “Broadband optical fibre with an attenuation lower than 0.1 decibel per kilometre,” Nat. Photon. 19, 1203 (2025).

# Supplementary Information

# Record-proof optical communication enabled by a quantum receiver with lossless optical decryption

## S1. Y-00 encryption process based on BPSK message modulation

Figure S1 illustrates the encryption process of the phase-shift-keying (PSK) Y-00 cipher system employed in this work. The inputs are a message bit sequence and a pre-shared secret seed key. The seed key, whose length $|K|$ is typically 256 bits, is expanded into a practically nonrepeated key stream by pseudorandom number generators (PRNGs) at the transmitter and receiver. For each symbol, $m$ bits are extracted from the key stream and converted into an encryption phase $\theta_{\mathrm{enc}}(k)$ through an irregular mapping function [S1]. In the present experiment, $m = 9$, resulting in $2^9 = 512$ possible phase values distributed over the interval $[-\pi/2, \pi/2)$. The irregular mapping is designed such that neighboring key-stream values are assigned to nonadjacent phase levels, making the effect of quantum noise on the key stream effectively random and thereby improving resistance against key-recovery attacks.

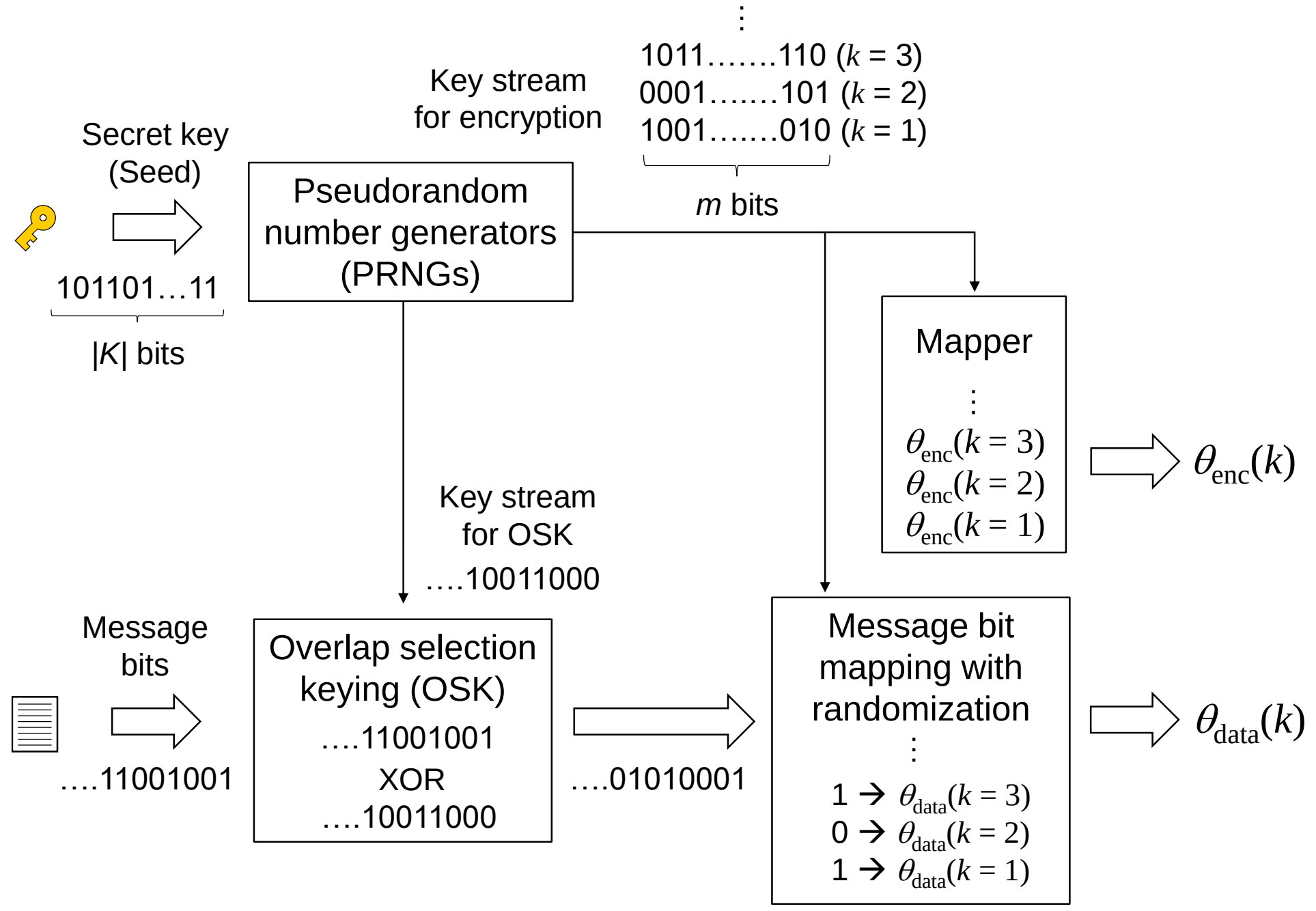

**Figure S1**: Encryption process based on Y-00 protocol. The encryption phase $\theta_{\mathrm{enc}}(k)$ and BPSK symbol phase $\theta_{\mathrm{data}}(k)$ are calculated from the secret seed key and message bits.

The message bits are first processed by overlap selection keying (OSK). Let $b(k)$ and $k_{\mathrm{OSK}}(k)$ denote a message bit and a key-stream bit assigned for OSK. The resulting bit is $b_{\mathrm{OSK}}(k) = b(k) \oplus k_{\mathrm{OSK}}(k)$, where $\oplus$ denotes the exclusive-OR (XOR) operation. By randomizing the relationship between the transmitted data and the original message-bit sequence, OSK enhances resistance against known-plaintext attacks (KPAs), in which an adversary attempts to recover the secret key using knowledge of portions of the transmitted data [S2]. Then, the least significant bit (LSB) of the key-stream segment for each encryption $k_{\mathrm{LSB}}(k)$ is additionally used for message-bit randomization. Specifically, $b_{\mathrm{r}}(k) = b_{\mathrm{OSK}}(k) \oplus k_{\mathrm{LSB}}(k)$ where $b_{\mathrm{r}}(k)$ denotes the randomized bit. The randomized bit is then mapped onto a BPSK symbol according to $\theta_{\mathrm{data}}(k) = \pi b_{\mathrm{r}}(k)$. This additional randomization causes the correspondence between the transmitted phase state and the message bit to vary according to the key stream on a symbol-by-symbol basis, preventing an unauthorized observer from directly inferring the message bit from coarse phase information such as the constellation quadrant or phase polarity.

The final transmitted phase is obtained by combining $\theta_{\mathrm{enc}}(k)$ and $\theta_{\mathrm{data}}(k)$. Therefore, each message bit is transmitted using one of $2^{m+1}$ ($2^{10} = 1024$ for $m = 9$) possible phase levels between 0 and $2\pi$ selected by the key stream. At the receiver, a legitimate receiver reproduces the identical key stream from the pre-shared secret key and determines $\theta_{\mathrm{enc}}(k)$ for each symbol. By compensating for the encryption phase in a symbol-by-symbol manner, the legitimate receiver recovers the original BPSK constellation and performs conventional binary demodulation. The impact of quantum noise on symbol discrimination is negligible under normal operating conditions. Subsequently, the message-bit randomization and OSK operations are removed using the corresponding randomization bits, $k_{\mathrm{LSB}}(k)$ and for OSK $k_{\mathrm{OSK}}(k)$, thereby recovering the original message bit. In contrast, an eavesdropper without knowledge of the key stream must discriminate among coherent states distributed over multiple possible phase levels while simultaneously lacking the correct bit-to-phase correspondence. Consequently, the signal discrimination error is significantly increased by quantum noise and measurement uncertainty, providing the physical-layer security foundation of the Y-00 protocol.

## S2. Theoretical bit error ratio analysis for homodyne, heterodyne, and optimal quantum receptions

We analyzed theoretical bit error ratios (BERs) of classical and quantum receivers in the manuscript. The BER curves of homodyne and heterodyne for different mean photon number per symbol $\bar{n}$ were calculated using the following formulas [S3]:

$$\mathrm{BER}_{\mathrm{homodyne}} = \frac{1}{2}\mathrm{erfc}\left(\sqrt{2\bar{n}}\right), \tag{S1}$$

$$\mathrm{BER}_{\mathrm{heterodyne}} = \frac{1}{2}\mathrm{erfc}\left(\sqrt{\bar{n}}\right). \tag{S2}$$

The BER of the optimal quantum receiver is given by Helström bound [S4]:

$$\mathrm{BER}_{\mathrm{Helstrom}} = \frac{1}{2}\left[1 - \sqrt{1 - \exp(-4\bar{n})}\right]. \tag{S3}$$

The Kennedy receiver is a near-optimal quantum receivers. According to Kennedy's original work [S5], the BER is given by

$$\mathrm{BER}_{\mathrm{Kennedy}} = \frac{1}{2}\exp(-4\bar{n}). \tag{S4}$$

## S3. Quantum-noise masking in PSK Y-00 cipher

Quantum-noise masking number is a primary security metric in a Y-00 cipher system. The number is defined as the number of signals masked by quantum noise and related to symbol error ratio (SER) when an eavesdropper discriminates encrypted high-order PSK signals. The quantum-noise masking number for a PSK Y-00 cipher was derived using semi-classical theory in [S6], where the mean photon number was assumed to be sufficiently large. Here, we derive the masking number in the absence of this assumption. Figure S2 shows the constellation diagram of high-order PSK signals encrypted based on the Y-00 protocol. We consider the case hat $M$-ary PSK is encrypted by phase randomization with $m$ bit resolution. The angle of adjacent encrypted signals $\Delta\theta_{\text{basis}}$ is obtained from the encrypted PSK order of $M \cdot 2^m$ as

$$\Delta\theta_{\text{basis}} = \frac{2\pi}{M \cdot 2^m}. \tag{S5}$$

The phase uncertainty imposed by shot noise $\Delta\phi_{\text{shot}}$ is calculated as

$$\sin\left(\frac{\Delta\phi_{\text{shot}}}{2}\right) = \frac{\sigma_{\text{shot}}}{A} \tag{S6}$$

where $\sigma_{\text{shot}}$ and $A$ are the standard deviation of shot noise and signal amplitude, respectively. The quantum-noise masking number $\Gamma_{\text{Q_psk}}$ is defined as the ratio of $\Delta\phi_{\text{shot}}$ and $\Delta\theta_{\text{basis}}$ and obtained as

$$\Gamma_{\text{Q_psk}} = \frac{\Delta\phi_{\text{shot}}}{\Delta\theta_{\text{basis}}} = \frac{M \cdot 2^m}{\pi}\sin^{-1}\left(\frac{\sigma_{\text{shot}}}{A}\right). \tag{S7}$$

An ideal attack for completely storing the encrypted high-order PSK signal is heterodyne detection that can simultaneously measure in-phase and quadrature-phase components. The variance of shot noise for the heterodyne is obtained as $\sigma_{\text{shot}}^2 = 1/2$. The signal amplitude $A$ has a relation of $A^2 = \bar{n}$, where $\bar{n}$ is the mean photon number per symbol. Thus, the masking number is expressed as

$$\Gamma_{\text{Q_psk}} = \frac{M \cdot 2^m}{\pi}\sin^{-1}\left(\frac{1}{\sqrt{2\bar{n}}}\right). \tag{S8}$$

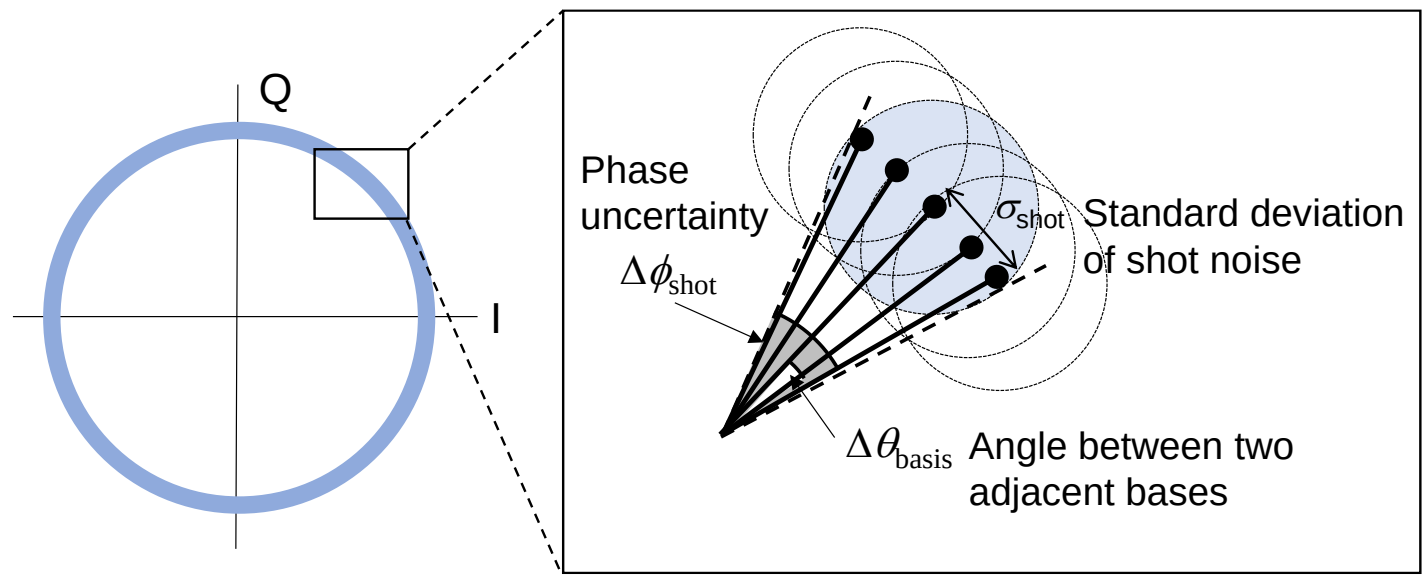

**Figure S2**: Constellation diagram of PSK Y-00 cipher.

Using Eq. (S8), we calculated the quantum-noise masking number $\Gamma_{Q_psk}$ for $2^{10}$ (=1024) PSK Y-00 cipher ($M$ = 2 and $m$ = 9) which are used in the experiments. Figure S3 shows the results. The theoretical sensitivity limit in the heterodyne to satisfy a pre-forward error correction (FEC) BER threshold of $1.81 \times 10^{-2}$, is calculated from Eq. (S2) to be 2.2 photon per symbol. Thus, the input signal power to the link should be 2.2 photon per symbol or less to achieve record-proof security. These results indicate that the quantum-noise making number is 162 at the input of the channel.

The quantum-noise masking number $\Gamma_{Q_psk}$ is related to the uncertainty of eavesdropper's detection of the encrypted PSK signal. One can estimate SER that an eavesdropper potentially reaches with ideal heterodyne reception. The SER is obtained from the masking number as

$$P_{\mathrm{SER_eve}} = erfc\left(\frac{1}{\sqrt{2}\Gamma_{\mathrm{Q_psk}}}\right) \tag{S9}$$

where *erfc* is the complementary error function [S6]. The SER for $\Gamma_{Q_psk}$ = 162 is higher than 0.995. Such a high SER implies that an eavesdropper obtains highly unreliable signal-level estimates, significantly reducing the effectiveness of subsequent key-recovery analysis.

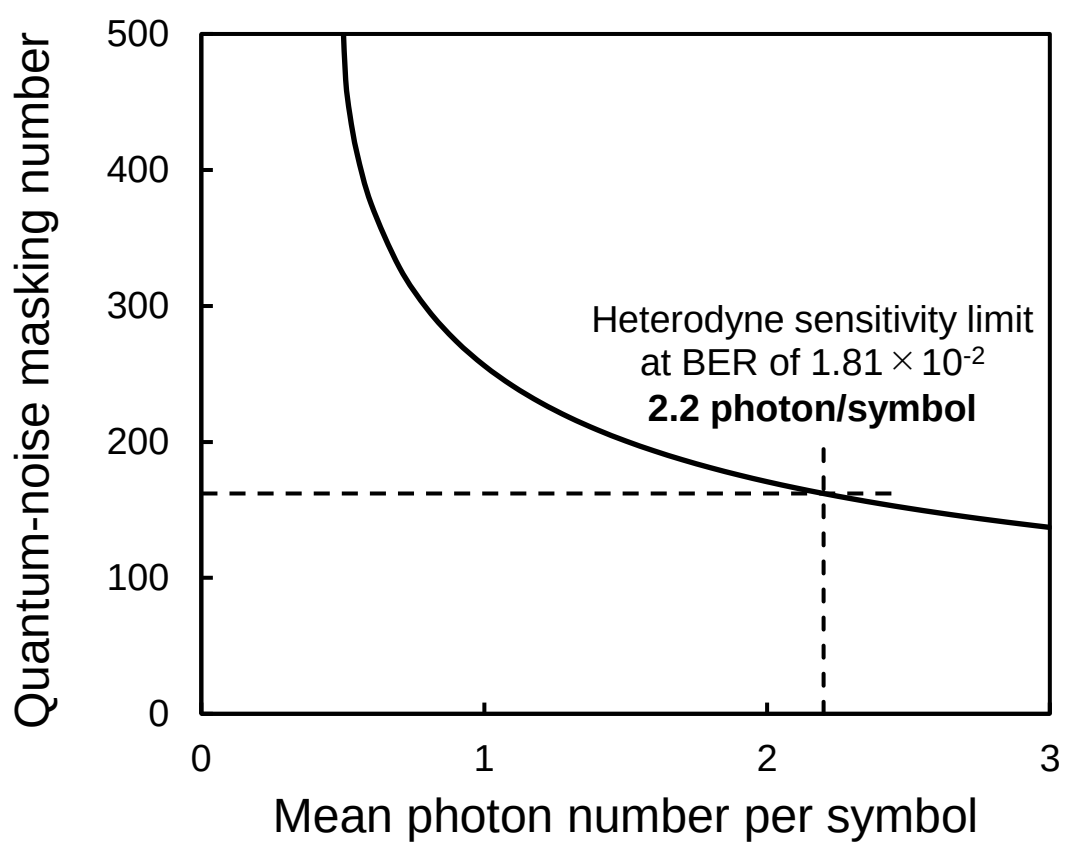

**Figure S3**: Quantum-noise masking number of 1024 ($M$ = 2, $m$ = 9) PSK Y-00 cipher for different mean photon number per symbol.

# References

1. T. Shimizu, O. Hirota, and Y. Nagasako, "Running key mapping in quantum stream cipher by Yuen 2000 protocol," Phys. Rev. A, 77(3), 034305, (2008).
2. O. Hirota, T. S. Usuda, and M. Fuse, "Quantum stream cipher Part III: design of keyed randomization and experiment", Proc. SPIE 5893, Quantum Communications and Quantum Imaging III, 589304 (2005).
3. K. Kikuchi, "Fundamentals of coherent optical fiber communications," J. Lightwave Technol. 34(1), 157–179 (2016).
4. C. W. Helstrom, "Quantum Detection and Estimation Theory," Academic Press, New York, (1976).
5. R. S. Kennedy, "A near-optimum receiver for the binary coherent state quantum channel," Research Laboratory of Electronics, MIT, Quarterly Progress Report, 108, 219-225 (1973).
6. K. Tanizawa, and F. Futami, "Ultra-long-haul digital coherent PSK Y-00 quantum stream cipher transmission system," Opt. Express, 29(7), 10451-10464 (2021).